\documentstyle[prb,aps,preprint]{revtex}
\begin{document}
\draft
\title{Bifurcations and a chaos strip in states of long
Josephson junctions}
\author{K.N. Yugay, N.V. Blinov, and I. V. Shirokov}
\address{Omsk State University, 55a Mira Ave., Omsk 644077, Russia}
\date{\today}
\maketitle
\begin{abstract}
Stationary and nonstationary, in particular, chaotic states in long
Josephson junctions are investigated.
Bifurcation lines on the parametric bias
current--external magnetic field plane are calculated.
The chaos strip along the bifurcation line
is observed. It is shown that
transitions between stationary states are the transitions from metastable
to stable states and that the thermodynamical Gibbs potential
of these stable states may be larger than for some metastable states.
The definition of a dynamical critical
magnetic field characterizing the stability of the stationary
states is given.
\end{abstract}

\section*{Introduction}

Dynamical chaos in long Josephson junctions is of great interest because
it can be a source of dynamical noise in devices based on them, in
particular, in SQUIDs, limiting the sensitivity of these devices.
Furthermore, dynamical chaos in long Josephson junctions (LJJ) is a very
interesting physical phenomenon taking place in nonlinear systems
in the absent of an external stochastic force.\cite{1,2,3,4,5,6,7,8,9}
Dynamical chaos in a LJJ is easily excited
and therefore it may also be investigated experimentally
rather easily. \cite{10,11}

In our previous works \cite{12,13} we have shown that among a set
of solutions of the Ferrell-Prange equation describing stationary
states of the LJJ in an external magnetic field\cite{14}
are both stable and unstable ones. At the same time, these stationary states
are asymptotic solutions of the nonstationary sine-Gordon equation
and we have also shown
that a selection of the stable solutions can be governed by a rapid
damping in time of the initial perturbation entering
into the nonstationary sine-Gordon
equation through the boundary conditions.
Changing the intensity of this perturbation at fixed shape,
we can obtain various stationary states for the LJJ without
a bias current or three clusters of states (stationary, and time dependent
regular and chaotic) in the presence of a bias current.
It turned out that asymptotic states are
very sensitive to an external perturbation, its value and shape define
the state (stationary, regular or chaotic)
to which the system will tend at $t\to \infty$ (we have called
this influence on the selection of asymptotic states
of the small rapidly damping initial perturbation in time
an effect of memory).
The fact of coexistence of all these three
characteristic asymptotic states selected only by the form
of the initial perturbation seems to be astonishing.
It is evidently enough
that the Ferrell-Prange equation will not have solutions at a large bias
current $\beta$. Therefore the question arises: at which  values of $\beta$
do stationary states of a LJJ disappear or what will be a boundary in the
parametric $\beta - H_0$ plane ($H_0$ is an external magnetic field)
that separates this plane on the regions where stationary states do
and do not exist?
Since the number of solutions of the Ferrell-Prange equation changes
at variation of the parameters $(H_0,\beta)$, another question arises:
what is the form of bifurcation lines in the plane $\beta - H_0$
that separate the parametric plane on the regions
with a different number of stationary
solutions of the Ferrell-Prange equation?

The existence of several stable solutions of the Ferrell-Prange equation
is equivalent to the fact that thermodynamical Gibbs potential
$G$ associated with the distribution of the magnetic field along
the junction has minima, and each minimum corresponds to
a certain solution of the Ferrell-Prange equation.
Does a global minimum of $G$ correspond to the most stable state
(e.g., in the Lyapunov sense)? In the case of the junction
of the finite length both Meissner and one-fluxon states
are thermodynamically advantageous simultaneously, so it is
interesting to investigate dynamical properties of these states.
Answering this question, we introduce a dynamiclal critical field that
describes the stability characteristic of the junctions.

In Sec.~1  bifurcation lines on the parametric $\beta-H_0$ plane are calculated.
In Sec.~2 the definition of the dynamical critical magnetic field is given
and the dependence of this field on $\beta$ and the length of the junction $L$
is calculated. In Sec.~3 transitions between states are described.
It is shown in Sec.~4 that a chaos strip arises along the bifurcation
line on the parametric $\beta-H_0$ plane.
The last Sec.~5 contains the discussion of our calculation and
brief conclusions.

\section{Bifurcation lines}

Stationary states of a LJJ are investigated using the numerical integration
of the Ferrell-Prange equation:
\begin{equation}
     \varphi_{xx}(x)=\sin\varphi(x)-\beta,
     \eqnum{1}\label{eq1}
\end{equation}
where $\varphi(x)$ is the stationary Josephson phase variable, $\beta$
is the dc bias current density normalized to the critical current $j_c$,
$x$ is the distance along the junction normalized to the Josephson
penetration length $\lambda_J=\sqrt{C\Phi_0 /8\pi^2 j_c d}$,
$\Phi_0$ is the flux quantum, $d=2\lambda_L+b$, $\lambda_L$ is the London
penetration length, $b$ is the thickness of the dielectric barrier.
The boundary conditions for Eq.\ (\ref{eq1}) have the form
\begin{equation}
     \varphi_{x}(x)\vert_{x=0}=
     \varphi_{x}(x)\vert_{x=L}= H_0,
     \eqnum{2}\label{eq2}
\end{equation}
where $L$ is the total length of the junction normalized to $\lambda_J$
and $H_0$ is the external magnetic field perpendicular to the junction and
normalized to $\tilde{H}=\Phi_0 / 2 {\pi} \lambda_J d$.

Numerical integration of Eqs.\ (\ref{eq1})--(\ref{eq2}) allows us to
find the regions with a certain number of solutions
on the parametric $\beta-H_0$ plane (Fig. \ref{fig1}).
It is easy to show that the set of points corresponding
to the even number of solutions forms two-dimensional domains
on this plane, whereas the set corresponding to
the odd ones may form just one-dimensional
curves.
Mostly, the lines corresponding to the odd number of the solutions
of the Ferrell-Prange boundary problem
coincide with the bifurcation lines.
Using the shooting method for solving of the boundary
problem one can prove that
the $2\pi$--periodicity of the function $H(\varphi_0)$
expressing the dependence of the magnetic field at the right
side of the junction ($x=L$) on the phase taken at the left side
($x=0$) results in the appearing of the odd number
of solutions only when the $H(\varphi_0)$ touches
the line $H=H_0$ in an extreme
point, i.e., $\partial H(\varphi_0)/\partial \varphi_0 =0$.
As an illustartion, we have plotted in Fig. \ref{fig2}
the function
$H(\varphi_0)$ at $H_0=0.5$, $L=5$, $\beta=0.25$ and $\beta=0.45$.

Boundaries between the regions -- bifurcation
lines -- define
an essential modification of the system. The bifurcation lines in Fig.1 are
obtained for $L=5$; here a step by $\beta$ is equal to $5\cdot 10^{-3}$
and a step by $H_0$ is equal to $2.5\cdot 10^{-3}$.
In this figure the numbers of solutions
of Eq.\ (\ref{eq1})--(\ref{eq2}) are pointed out,
the numbers of  stable solutions are given in the
brackets, while  $M$ and $1f$ denote a stable Meissner and one-fluxon
states, respectively. It is seen that a Meissner state is stable
at small values of $H_0$
and at large values of $H_0$ a one fluxon state is stable.
It should be noted that the region where there are no stationary solutions
(region 0) bounds  with  the  region  having  a  minimum  of  stationary
solutions,
being equal to 2 (region 2). In approaching the boundary of region 0
and 2 the number of stationary solutions decreases:
$6\to4\to2\to0$, on the other hand, a number of nonstationary states
which are
the asymptotic solutions of the sine-Gordon equation, increases.
Our calculations have shown that one of two stationary solutions in region 2
is stable,  and another is unstable (metastable).
We noted earlier \cite{12} that the stable states are symmetrical.
The presence of bias current $\beta$ leads to a symmetry violation
that results, evidently, in the instability of the states.

The problem of the stability of stationary
states $\varphi(x)$ was solved in the following way: \cite{13}
the sine-Gordon equation was linearized in the vicinity of stationary
solution: $\varphi(x,t) = \varphi(x) + \theta(x,t)$, where $\theta(x,t)$
is the infinitesimal perturbation. The equation for $\theta(x,t)$ -- the
linearized sine-Gordon equation -- we can solve by means of the expansion
of this function in terms of a complete system of
eigenfunctions of the Schr\"odinger operator with potential
$\cos\left[\varphi(x)\right]$:
\begin{equation}
   \theta(x,t) = \sum_n e^{\lambda_n t}u_n(x), \eqnum{3}    \label{eq3}
\end{equation}
where $u_n(x)$ are eigenfunctions of the  Schr\"odinger operator
of the problem:
\begin{equation}
-u_{xx}(x) + u(x) \cos \varphi(x) = E u(x),    \eqnum{4}    \label{eq4}
\end{equation}
\[
   u_x(x)\vert_{x=0}= u_x(x)\vert_{x=L} = 0 ,
\]
and
\begin{equation}
     \lambda_n = -\gamma \pm \sqrt{\gamma^2 - E_n},   \eqnum{5}   \label{eq5}
\end{equation}
where $\gamma$ is the dissipative coefficient in the sine-Gordon equation.
We note that values of $\lambda$ coincide with
corresponding values of Lyapunov exponents in the case when perturbations
are considered with respect to the stationary solutions. In general case,
Lyapunov exponents are calculated in the same way as in Ref. 13.
Thus, in the presence of a bias current we have the different picture
of a LJJ states than at $\beta=0$ (this case
has been examined in Ref. 12).
For example, at $H_0=1.9$ the increasing of $\beta$ from 0 to 0.22
leads to the changing of the stationary states number
$6\to4\to2\to0$, i.e., to a consecutive losing
of the stationary solutions. Simultaneously, an increasing of the number
of nonstationary states occurs that we found by directly solving
the nonstationary sine-Gordon equation.

\section{Dynamic critical field}

In the literature the critical magnetic field $H_{c_1}$ in a LJJ
is defined as a field value, at which an existence of a Josephson vortex 
(fluxon, soliton) becomes
advantageous thermodynamically for the first time
(see, for example, Refs. 10 and 11).
In the case of an infinitely long junction the critical field is
$
H_{c_1}(\infty)=4/\pi \simeq 1.274.
$
Essentially, this field corresponds to the global
minimum of the thermodynamic Gibbs potential for the one-fluxon state.
However, in a junction of finite length there are some local minima
that coexist with the global one and every minimum corresponds
to the solution of Eqs. (\ref{eq1})--(\ref{eq2}). Some of these
solutions are stable, another unstable in the sense discussed in
Sec.~1.

We write down the thermodynamic Gibbs potential in the form
\begin{equation}
G=\int_{0}^{L}dx[{1\over 2}{\varphi_x}^{2}(x)+1-\cos \varphi(x)-
   \beta \varphi(x)-H_0{\varphi_x}(x)].
   \eqnum{6}  \label{eq6}
\end{equation}
Here $G$ is the thermodynamic Gibbs potential per unit length along
an external magnetic field and normalized to
$\tilde{G}=\Phi_0 / 16 \pi^3 \lambda_{J}d$.
The Ferrell-Prange equation is an extremal of the
functional (\ref{eq6}).
An investigation of the second variation of $G$ shows that all extrema
of this functional satisfy to the necessary and sufficient conditions of
strong minimum. \cite{15} Thus, all solutions of Eqs.\ (\ref{eq1})--(\ref{eq2})
(both stable and unstable ones) correspond to minima
of the thermodynamic Gibbs potential;
one of them is global, the others are local. Our calculations
of the thermodynamic Gibbs potential (\ref{eq6}) show that, for
example, at $\beta=0$, $L=5$ and $H_0=0.67$ the Meissner state
has a global minimum ($G_M=-0.44$), but the stable one-fluxon
state has a local one ($G_{1f}=4.03$). The one-fluxon state
has a global minimum of $G$ starting at
$H_0=1.57$ ($G_{1f}=-2.582$) and at the same value of $\beta$ and $L$.
At this value of $H_0$ a Miessner state
has a local minimum $G_M=-2.58$. At $H_0\ge 2.09$ the Meissner state
disappears. Thus, at a field less than the critical one $H_{c1}$,
the stable one-fluxon state exists. We shall further call the minimum
value of a magnetic field at given $L$ and $\beta$, at which the stable
one-fluxon state appeares for the first time and which corresponds to the
local minimum of the thermodynamics Gibbs potential as the dynamical
critical field $H_{dc}$.
It is interesting that the dynamical critical field $H_{dc}$
makes up on the parametric plane a line that coincides with the bifurcation
line BC (see Fig. \ref{fig1}).  Our calculations show that the bias
current increases the dynamical critical field $H_{dc}$. Evidently,
it is connected with a symmetry violation of a state by the bias
current $\beta$. In Fig.\ \ref{fig3} two stable one-fluxon states
at $\beta=0$ and $\beta=0.1$ ($L=5$, $H_0=1.4$) are shown.
It is seen that the state with $\beta=0.1$ is asymmetric.
The dynamical critical field at $L=5$ are $H_{dc}=0.67$
at $\beta=0$ and $H_{dc}=1.4$ at $\beta=0.1$. Upon increasing
$L$ the value of $H_{dc}$ is changed
($\beta=0$): $H_{dc}(L=5)=0.66$, $H_{dc}(6)=0.4$, $H_{dc}(7)=0.26$,
$H_{dc}(8)=0.15$, $H_{dc}(10)=0.06$, i.e. the $H_{dc}$
decreases. In this case the critical field $H_{c1}$ has the values:
$H_{c1}(L=5)=1.57$, $H_{c1}(6)=1.45$, $H_{c1}(7)=1.38$,
$H_{c1}(8)=1.34$, $H_{c1}(10)=1.28$, i.e., the $H_{c1}$
decreases also approaching to the value of
$H_{c1}(\infty)\simeq 1.274$.

\section{Transitions between states}

As it has been shown in the previous section,
every stationary state of LJJ, i.e., the solution
of Eqs.\ (\ref{eq1})--(\ref{eq2}),  corresponds to a minimum
of the thermodynamic Gibbs potential and these minima are not equivalent
with respect to the problem of instability.
For example, in Fig.\ \ref{fig4} stationary states of LJJ
at $H_0=2.035$, $\beta=0.001$
and L=5 are shown. The values of the Gibbs potential
calculated using Eq. (\ref{eq6})
are as follows: $G_1=-5.03, G_2=-4.52, G_3=-4.61, G_4=-4.64$,
$G_5=-4.61, G_6=-6.7$. States 4 (Meissner) and 6 (one-fluxon) are stable,
the other ones are metastable. It should be noted that unstable
state 1 corresponds to
deeper minimum than the stable state 4. This property contradicts
the naive idea that more stable states occur at deeper
minima. Now we shall consider this question in detail.

The sine-Gordon equation with dissipation and bias current describing
an evolution of initial state has the form:
\begin{equation}
  \varphi_{tt}(x,t)+2\gamma\varphi_t (x,t) -
  \varphi_{xx}(x,t)=-\sin\varphi(x,t)+\beta,       \eqnum{7}    \label{eq7}
\end{equation}
where $t$ is a time normalized to the inverse of the Josephson plasma
frequency
$
  \omega_J=\sqrt{2\pi c j_c/ C\Phi_0},
$
$C$ is the junction capacitance per unit area,
$
\gamma=\Phi_0 \omega_J / 4 \pi cRj_c
$
is the dissipative coefficient per unit area,
$R$ is the resistance of junction per unit area. We write down the boundary
conditions for Eq.\ (\ref{eq7}) in the form
\begin{equation}
     \varphi_{x}(x,t)\vert_{x=0}\equiv
      H(0,t) =
     \varphi_{x}(x,t)\vert_{x=L}
\equiv H(L,t)=H_0\left(1-a e^{-t/2t_0}\cos 0.5t\right). \eqnum{8} \label{eq8}
\end{equation}

The integration of Eqs.\ (\ref{eq7})--(\ref{eq8})
for $H_0=2.035, \beta=0.001$,  $L=5$
(the same as in Fig.\ \ref{fig4}) and $\gamma=0.26$
gives: the metastable state 1 passes to the stable
state 6 at any values of perturbation parameter $a$, $2\to 4$ at $a$=0,
$2\to 6$ at $a$=1, $3\to 4$ at $a$=0.05, $3\to 6$ at $a$=0.07, $4\to6$
at $a$=0.5 and so on. Every transition from the metastable state to the
stable one, $m\to n$, is a transition from the state with the certain value of
local minimum $G_m$ to another state with smaller value of minimum $G_n$.
These transitions $m \to n$ with $G_m>G_n$ are realized by certain values
of the parameter of the initial perturbation $a$ in expression (\ref{eq8}).
One can say that the local minima of $G_l$ are connected with each other by
a certain disintegration channel along the coordinate $a$.
From this point of view one can say also that stationary
states contain a specific ``latent'' parameter, by which a connection with
different local minima $G_l$ may be realized. In particular,
the perturbation parameter $a$ appears here as a ``latent''
parameter. It is possible, there are several ``latent'' parameters
connecting the stationary states. One of the most important characteristics
of ``latent'' parameters is that the stationary state
does not depend on them
directly; however, the form of the asymptotic state and
the rate of disintegration
depend essentially on them. The presence of a ``latent'' parameter
apparently explains, a nonequivalence of the different local
minima with respect to the stability, especially in the case
when a stable local
minimum is above of a nonstable local one.
In Table\ \ref{tab1} results of the integration of
Eqs.\ (\ref{eq1}), (\ref{eq2}) and the calculation
of $G$ for every of these solutions at $H_0=1.174, \beta=0, L=8$
are represented.
The transitions between states $m\to n$ are defined as follows:
the {\it m}th solution of the stationary Ferrell-Prange
equation (\ref{eq1})--(\ref{eq2})
was taken as an initial condition of the sine-Gordon
equations (\ref{eq7}) and (\ref{eq8}).
If this {\it m}th state was unstable then it  fell
into the {\it n}th stable state.

The scheme of the transitions between states $m\to n$ is
represented in Fig.\ \ref{fig5}.
It is seen, that $G_m > G_n$ for all the transitions
(we note that $G_3$ and $G_9$ for the metastable states 3 and 9 are less
than $G_{12}$; the state 12 is stable). The stable
states -- Meissner, one-fluxon, and two-fluxon --- are shown
in Fig.\ \ref{fig6} at the same parameters as in Fig.\ \ref{fig5}.

\section{Chaos strip}

As we noted above, a number of stationary states decreases with
approaching to the bifurcation line 0 - 2, but the number of
nonstationary asymptotic states is increased simultaneously.
Changing the perturbation parameter $a$ we can obtain three sorts of typical
states: stationary, regularly and chaotic. \cite{13}
These states  are distinguished
not only by a form of the field distribution in the junction and a variation
in time, but also by values of the Lyapunov exponent $\lambda$:
for the stationary states $\lambda < 0$, for the regular states
$\lambda \leq 0$ and for the chaos states $\lambda >  0$.
The Lyapunov exponents were calculated in the same way as in Ref. 13.
However, as the calculations have shown, chaotic states may be excited not
in the whole region 2 (see Fig.\ \ref{fig1}),
but only in the bounded region in close
to the bifurcation line 0 -- 2. This region is extended in the form of a narrow
strip along the bifurcation line 0 -- 2 approximately from 0.7 to 1.6
in $H_0$ and in the range of 0.002 -- 0.015 in $\beta$.
We note, that the chaos strip
is arranged mostly under the bifurcation line in the region 2,
but not in the region 0, as it may be expected because of all states
in the region 0 are nonstationary.
The chaos strip is outlined on the parametric
$\beta - H_0$ plane in Fig.\ \ref{fig1}.

This chaos strip along the bifurcation line 0 -- 2 calls to mind (to a
certain extent) the separatrix of a nonlinear oscillator, where a chaos
motion is observed.

\section{Discussion and conclusions}

In the present work we have shown that the parametric $\beta - H_0$ plane
of a LJJ is separated on series of regions with the different number
of solutions of the stationary Ferrell-Prange equation. The boundaries
between these regions---bifurcation lines---characterize an essential
modification of the system. A chaos strip arises
along the bifurcation line 0 - 2.  We have found that the chaos
strip is arranged in the main
below the bifurcation line 0 - 2, where stationary states take place.

We have introduced the definition of a dynamical
critical field as the lowest field at which
the one-fluxon state becomes stable for the first time
in the Lyapunov's sense.
In addition, the Meissner state
may also be stable at same parameters. Because both the Meissner and the
one-fluxon states may be thermodynamically advantageous simultaneously,
our definition based on the stability in the Lyapunov's sense characterizes
an important feature of the stationary states of the LJJ.

We have shown that disintegration of the metastable states and the transition
to some stable states $m\to n$ occur for $G_m > G_n$.
A metastable state corresponds to the local minimum of the Gibbs potential,
and also this minimum may be lower than this one
of a stable state. A nonequivalence of these local minima we
explain by means of existing of a ``latent'' parameter not detecting
in a stationary state, by which, for example,
two local minima may be connected
and a channel of the disintegration of the upper state may arised.
In our case the perturbation parameter plays a role of the a ``latent''
parameter, however, the number of these parameters may be much greater.
We note the anology between the quantum transitions and the transitions
mentioned above, although the system is described by the classical
Ferrell-Prange and sine-Gordon equations.

We are aware that we could not touch upon all questions concerning
the properties of a LJJ.
We hope to return to the problems of a LJJ in our next work.

These investigations are supported by the Russian Foundation for
Basic Research (project N. 96-02-19321).

\begin{figure}
\caption{Bifurcation lines.
The numbers of solution of the Ferrell-Prange equation (1)--(2)
are pointed out. The number of stable states is
indicated in brackets. $M$ denotes a stable Meissner state and $1f$ denotes
a stable one-fluxon state. $L = 5$.}
\label{fig1}
\end{figure}

\begin{figure}
\caption{Dependence of the magnetic field at $x=L$
on the phase taken at the left side of junction $x=0$
at $H_0=0.5$, $L=5$, $\beta=0.25$ and $\beta=0.45$.}
\label{fig2}
\end{figure}

\begin{figure}
\caption{One-fluxon states at $H_0=1.4$ and $L = 5$
for $\beta = 0$ and $\beta = 0.1$.}
\label{fig3}
\end{figure}

\begin{figure}
\caption{Stationary states of LJJ at $H_0 = 2.035$, $\beta = 0.001$
and $L = 5$. States 1, 2, 3, 5 are unstable, states 4 and 6 are stable.}
\label{fig4}
\end{figure}

\begin{figure}
\caption{The scheme of transitions between states $m\to n$.
States 8, 10 and 12 are stable (8---Meissner, 10---one-fluxon,
12---two fluxon),
others are unstable. $H_0=1.174$, $L=8$, $\beta=0$.}
\label{fig5}
\end{figure}

\begin{figure}
\caption{The stable states: $M$---Meissner, $1f$---one-fluxon,
and $2f$---two-fluxon at the same parameters as those in Fig. 5.}
\label{fig6}
\end{figure}

\begin{table}
\begin{tabular}{dcccc}
Number of state&Stability&$G$&Transitions $m\to n$&Sort of stable states \cr
\hline
1 & unstable & 2.34 & $1\to 10$ & \cr
2 & unstable & 2.78 & $2\to 8$ & \cr
3 & unstable & 0.64 & $3\to 8$ & \cr
4 & unstable & 14.69 & $4\to 10$ & \cr
5 & unstable & 14.98 & $5\to 10$ & \cr
6 & unstable & 14.69& $6\to 10$ & \cr
7 & unstable & 13.53 & $7\to 10$ & \cr
8 & stable & -1.42 & $8\to 8$ & Meissner \cr
9 & unstable & 0.64 & $9\to 8$ & \cr
10 & stable & -0.44 & $10\to 10$ & 1 fluxon \cr
11 & unstable & 2.34 & $11\to 10$ & \cr
12 & stable & 2.29 & $12\to 12$ & 2 fluxon \cr
\end{tabular}
\caption{Transitions between states}
\label{tab1}
\end{table}


\begin{references}
\bibitem{1} W.J. Yeh, O.G. Symko, and D.J. Zheng, Phys. Rev. {\bf B 42},
        4080(1990).
\bibitem{2} N. Gronbech-Jensen, P.S. Lomdahl, and M.R. Samuelsen, Phys. Rev.
        {\bf B 43}, 1279(1991).
\bibitem{3} N. Gronbech-Jensen, Phys. Rev. {\bf B 45}, 7315(1992).
\bibitem{4} G.F. Eriksen and J.B. Hansen, Phys. Rev. {\bf B 41}, 4189(1990).
\bibitem{5} X. Yao, J.Z. Wu, and C.S. Ting, Phys. Rev. {\bf B 42}, 244(1990).
\bibitem{6} L.E. Guerrero and M. Ostavio, Physica {\bf B 165 - 166},
          11657(1990).
\bibitem{7} I.H.Dalsgaard, A. Larsen, and Mygind, Physica {\bf B 165 - 166},
        1661(1990).
\bibitem{8} S. Rajasekar and M. Lakshmanan, Physica {\bf A 167}, 793(1990).
\bibitem{9} S. Rajasekar and M. Lakshmanan, Phys. Lett. {\bf A 147}, 264(1990).
\bibitem{10} A. Barone and G. Paterno,  Physics and Applications
        of the Josephson Effect (Wiley-Interscience, New-York, 1982).
\bibitem{11} K.K. Likharev, Introduction to Dynamics of Josephson
        Junctions (Nauka, Moscow, 1985) (in Russian).
\bibitem{12} K.N. Yugay, N.V. Blinov, and I.V. Shirokov, Phys. Rev. {\bf B 49},
        12036(1994).
\bibitem{13} K.N. Yugay, N.V. Blinov, and I.V. Shirokov, Phys. Rev.
        {\bf B 51}, 12737(1995).
\bibitem{14} C.S. Owen and D.J. Scalapino, Phys. Rev. {\bf 164}, 538(1967).
\bibitem{15} I.M. Gelfand and S.V. Fomin, Calculus of Variations
         (Fizmatgiz, Moscow, 1961) (in Russian).
\end{references}
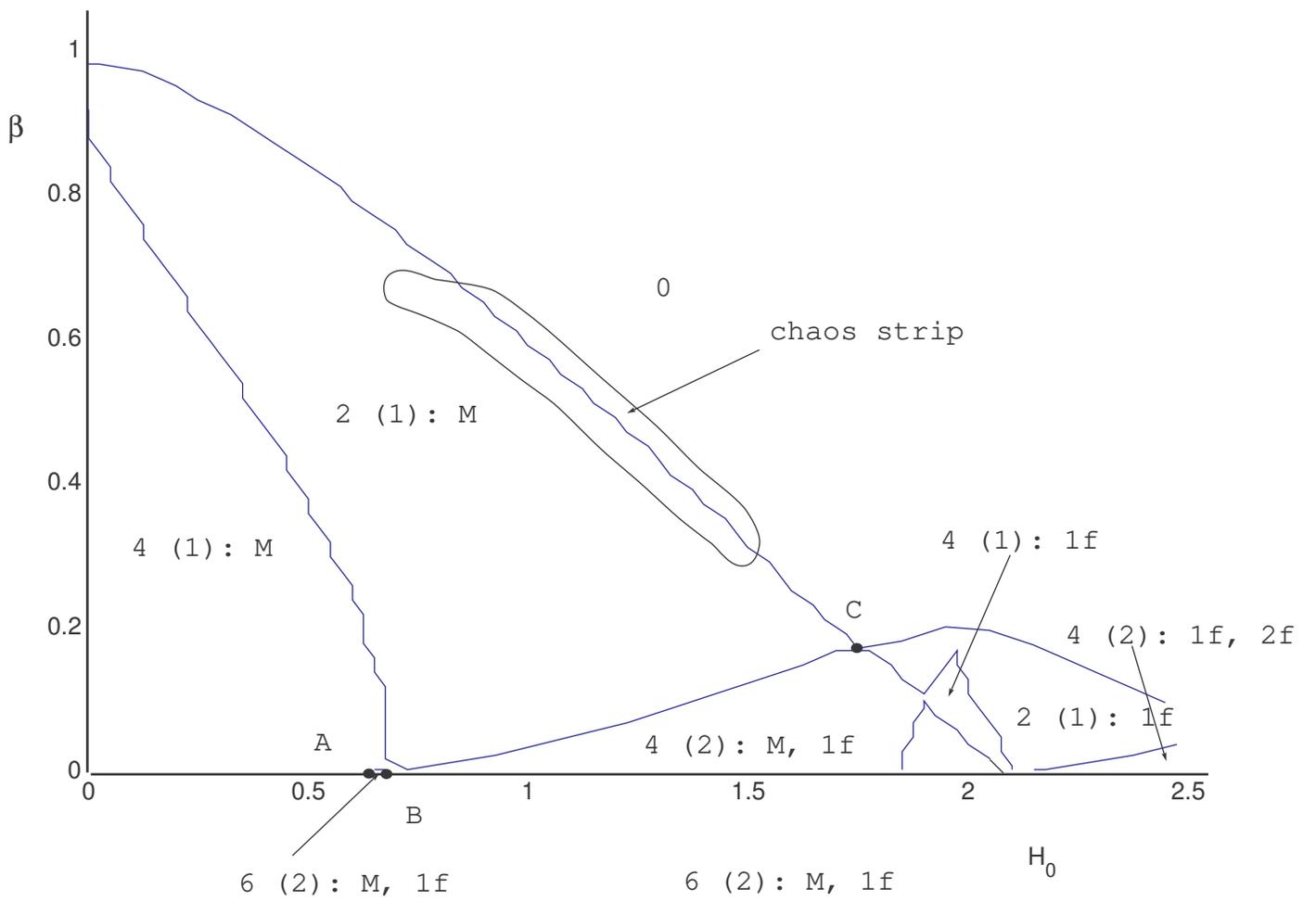
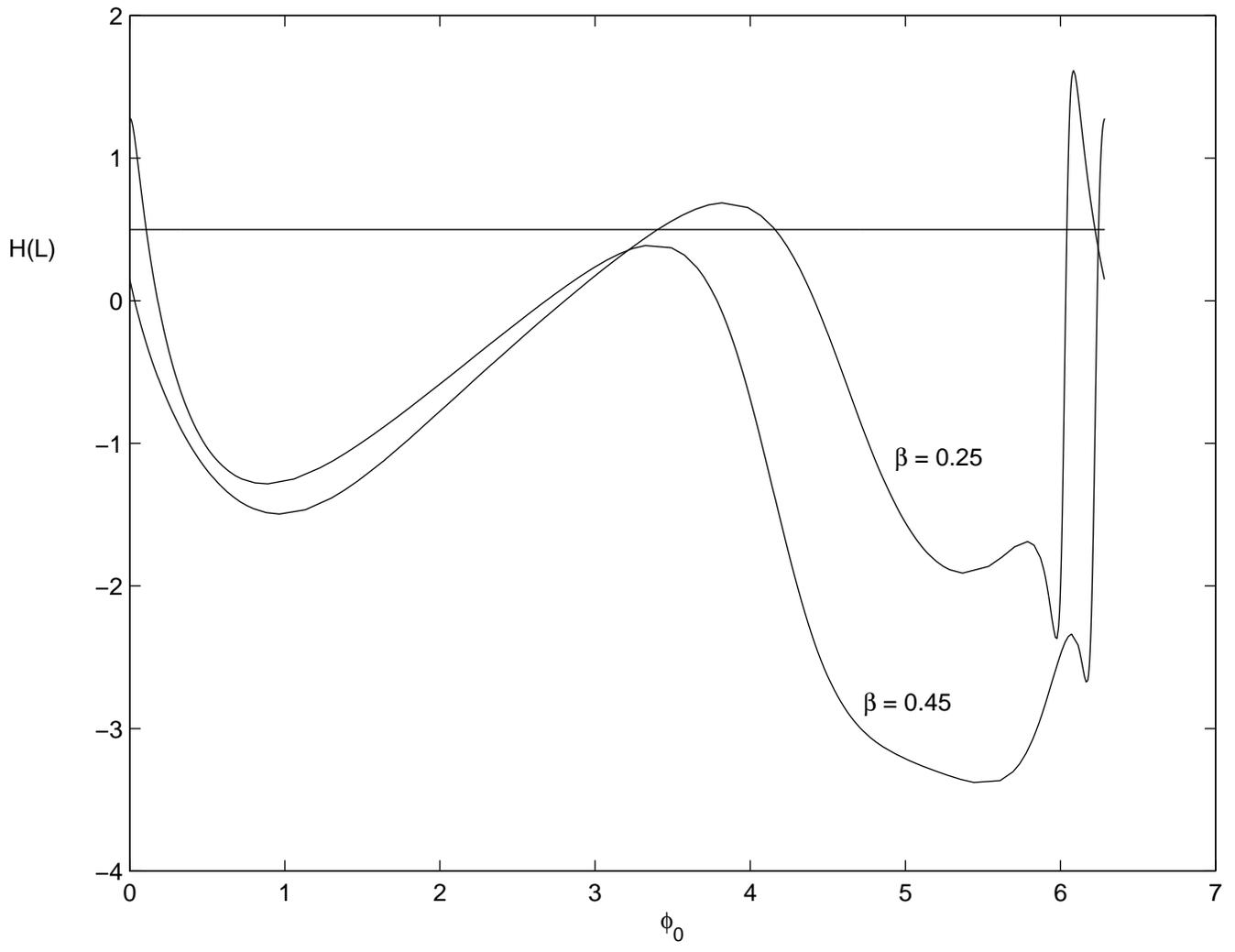
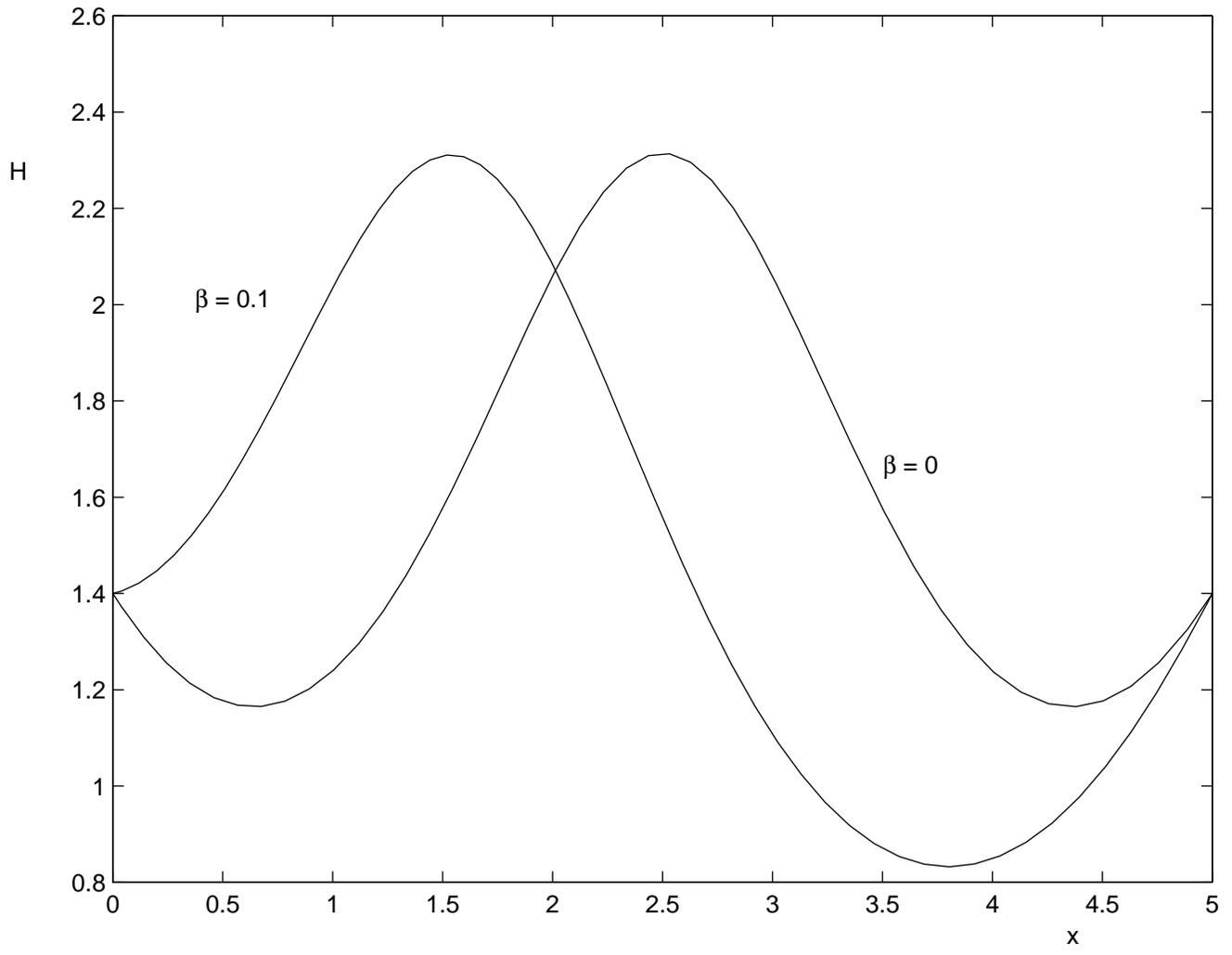
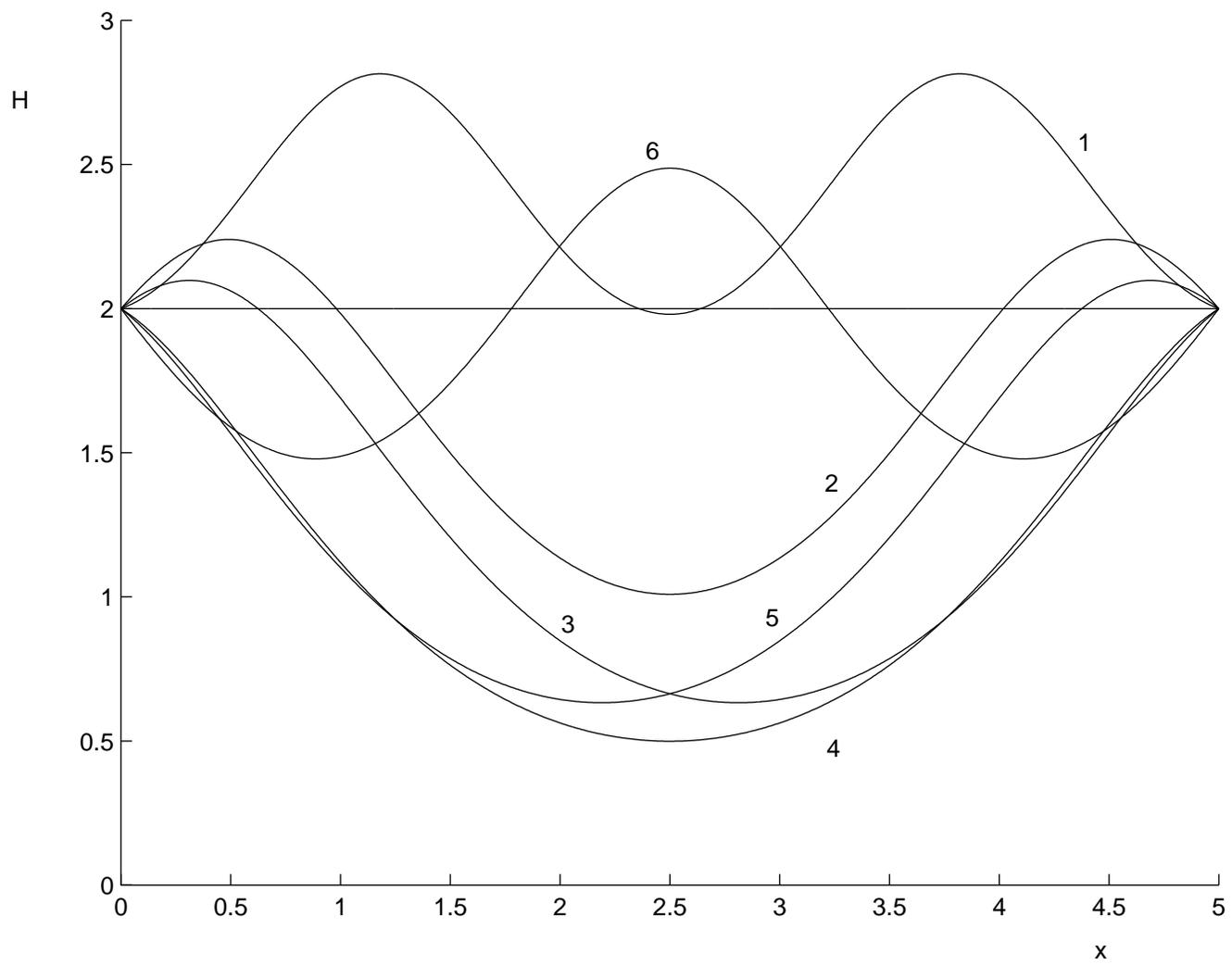
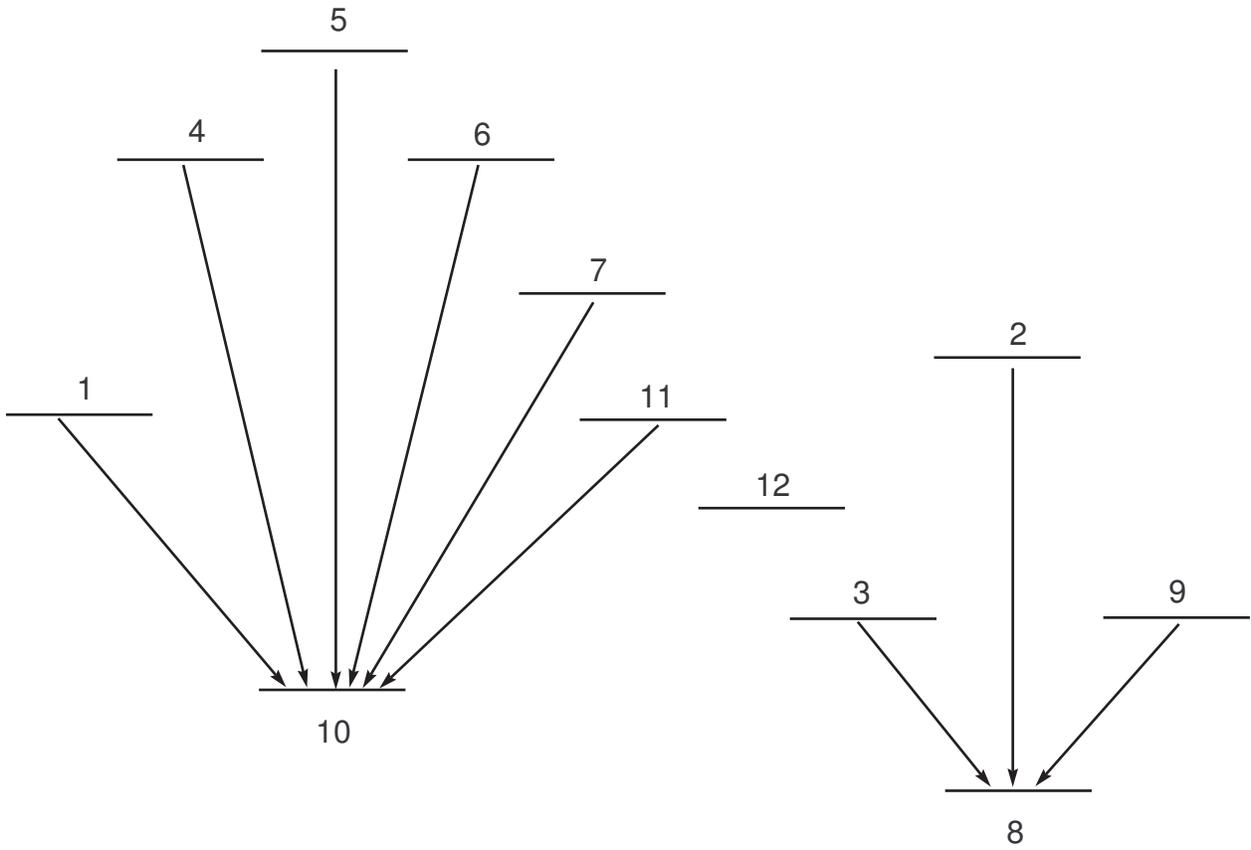
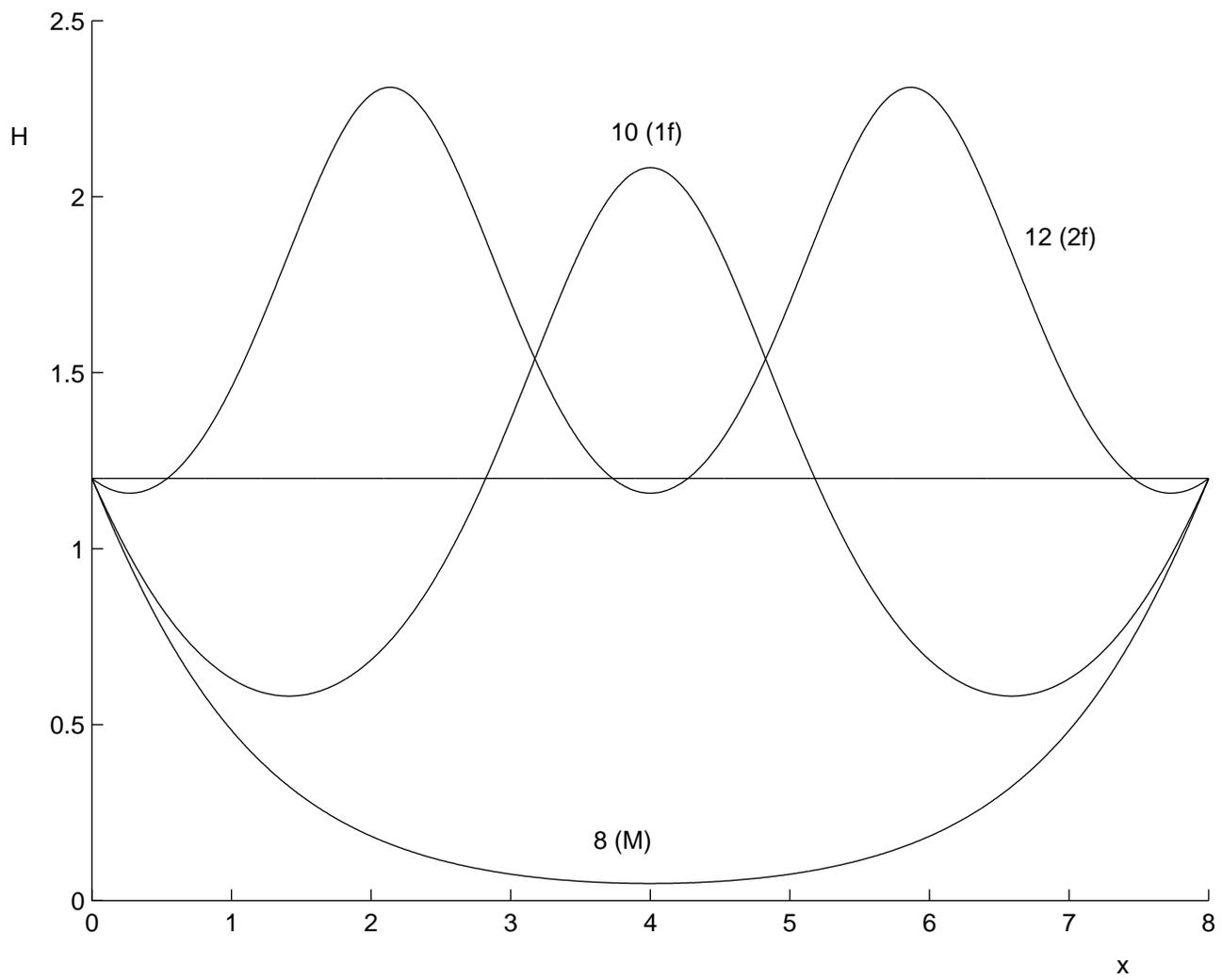
\end{document}